\documentstyle[editedvolume,namedreferences]{crckapb}

\def\mathrelfun#1#2{\lower3.6pt\vbox{\baselineskip0pt\lineskip.9pt
  \ialign{$\mathsurround=0pt#1\hfil##\hfil$\crcr#2\crcr\sim\crcr}}}
\def\simlt{\mathrel{\mathpalette\mathrelfun <}}

\def\s{\scriptscriptstyle}
\def\rc{{r_{\rm c}}}
\def\rt{{r_{\rm t}}}
\def\alphac{{\alpha_{\rm c}}}
\def\alphat{{\alpha_{\rm t}}}

\begin{opening}
\title{WEAK LENSING AND THE SLOAN DIGITAL SKY SURVEY}
\subtitle{Wide Area Weak Lensing}

\author{Albert Stebbins}
\author{Tim Mckay,}
\author{Joshua A. Frieman}
\institute{Fermi National Accelerator Laboratory\\
           Box 500, Batavia IL 60510, USA}

\end{opening}

\runningtitle{WEAK LENSING AND THE SDSS}

\begin{document}

\begin{abstract}
While the strategy for the first applications of weak lensing has been
to ``go deep'' it is equally interesting to use one's telescope time to instead
``go wide''.  The Sloan Survey (SDSS) provides a natural framework for a very
wide area weak lensing survey.
\end{abstract}

Probing of the mass distribution using the distortion of galaxy image shapes by
the intervening gravitational field this mass produces is a powerful new
technique for probing cosmological structure \cite{VTJ,TVW,ME,BSBV,K92}.  The
``weak lensing'' technique will no doubt become one of the standard probes, on
par with galaxy redshift surveys and maps of CMBR anisotropies.  Except for
small areas on the sky near distant rich clusters or very near galaxies, the
image distortion is expected to be small and weak lensing is the appropriate
technique. One must average over many galaxies to obtain a significant
detection of the small image distortion; typically by measuring correlations in
galaxy position-angles and thus the shear.  Deep imaging is extremely useful as
it allows one to get accurate estimates of the shapes of large numbers of
background galaxies in the relatively small field of view of most
telescopes. If one fails to go deep one can identify fewer background galaxies
and, in any case, one obtains only accurate shape information for the brighter,
larger galaxies.  However even with moderately deep images one can, in
principle, use the weak lensing technique to infer the foreground mass
distributions.  If the the number of galaxies per unit area for which one has
accurate shape information is small then one should survey a larger area to
obtain a significant signal.

	The first successful applications of the weak lensing
\cite{TVW,SEF94,FKSW} has naturally been to take deep images of galaxies behind
rich clusters where the shear is large, and perhaps more importantly, where one
has a fairly good idea what one expects to find. Attempts have also been made
to detect shear in the field (i.e. a direction not associated with a particular
galaxy concentration), but without any definitive detection \cite{MBVBSSK}.
When looking at the field one can expect to find contributions to image
distortions from mass at various distances along the line-of-sight.  While it
would be useful to study the statistical properties of the shear at a given
depth, one will, in the end, want to chart how the shear varies with depth.  By
understanding the variation with depth one can learn about the radial
distribution of densities along different lines-of-sight.  This can tell us
something about the evolution of the density field and in particular about
cosmological parameters such as $\Omega$ and $\Lambda$; as well as allow one to
construct a crude map of the mass distribution.  The latter application is
particularly interesting as it will allow one to compare the mass distribution
with the better studied {\it nearby} galaxy distribution \cite{VG94}.  Thus
even if one had a very deep survey of galaxy image shapes one would want to
study the dependence of shear with depth and in effect look at less deep
surveys.  The study of shear at $z\sim0.1-0.4$ is interesting in it's own
right!

	Any imaging survey of the sky is implicitly measuring the shapes of the
galaxies it is able to detect.  As long as the combination of depth and area of
the survey are large enough to obtain a sufficient S/N one can in principle use
this for weak lensing.  One's calculation of depth must take into account the
accuracy with which one is able measure the galaxy shapes.  However it is
generally true that one does not loose much by even relatively large random
errors in the galaxy shapes.  This is because the intrinsic non-circularness of
the true projected galaxy shapes introduces random uncertainties in the
inferred shear and one would have to make fairly large measurement errors to
significantly add to these uncertainties.  The true galactic position-angles
are (assumed) random and therefore by using a sufficiently large number of
galaxies one can reduce both the intrinsic and measurement uncertainties if
they are random.  However if uncorrected measurement errors are correlated
between different galaxies one may never reach an acceptable S/N.  Since the
deeper the survey is the larger the signal will be, the requirement to control
these systematic errors is less.  It is not clear to what level one can reduce
systematic errors and it is thus not clear how shallow a survey one could use
for weak lensing studies.

	The Sloan Digital Sky Survey (SDSS) \cite{Kent94} is a prime example of
a large imaging survey on which one may ``piggy-back'' a weak lensing program
\cite{VG94}.  Perhaps the most publicized aspect of the SDSS is a redshift
survey of $10^6$ galaxies.  To obtain the redshift targets the SDSS will image
1/4 of the sky in 5 colors, mostly around the North Galactic Cap, identifying
galaxies in the North down to a nominal magnitude limit of $r'<23.1$, and going
to 25.1 in parts of the Southern survey.  This will yield a catalog of
$\sim5\times10^7$ galaxy images.  One does not really need the galaxies near
the limiting magnitude to obtain a significant weak lensing signal and in this
sense the SDSS can expect to do much better than a marginal detection.  The
multi-color photometry will be extremely useful for weak lensing as we expect
to determine galaxy redshifts to $\Delta z\sim0.04$ {\it photometrically}
\cite{Szalay}.  With this redshift information one can map the shear as a
function of distance. This allows one to better localize the mass distribution
as a function of radius and make more of a direct comparison of the mass and
galaxy distributions.  Of course the SDSS redshift survey gives exactly the
galaxy distribution one would want to compare to the mass distribution
determined via weak lensing from the imaging survey.

	In Gould and Villumsen  (1994) it was estimated that nearby clusters
and their  extended halos would dominate the shear field measured by the SDSS.
To illustrate some of the above comments in this regard let us consider the
mean shear given by a model cluster with radial density profile
\begin{equation}
\rho(r)={3v_{\s1}^2\rt^2\over2\pi\,G(r^2+\rc^2)(r^2+\rt^2)} \qquad \rt\ge\rc
\end{equation}
which is a kind of truncated non-singular isothermal sphere.  For this profile
the image shear as a function of angle, $\alpha$, from the cluster center is
\begin{equation}
\gamma(\alpha)=6\pi{v_{\s1}^2\over c^2}{\beta\over\alpha}
      {\alphat^2\over\alphat^2-\alphac^2}
      \left(2{\alphat-\alphac\over\alpha}
            -{\alpha^2+2\alphat^2\over\alpha\sqrt{\alpha^2+\alphat^2}}
            +{\alpha^2+2\alphac^2\over\alpha\sqrt{\alpha^2+\alphac^2}}\right).
\end{equation}
Here $\alphac$ and $\alphat$ are the angles subtended by $\rc$ and $\rt$ at the
distance of the cluster.  The distances of the galaxies whose shear is measured
comes into the factor $\beta$.  If most of galaxies are much further away than
the cluster then $\beta\approx1$ while if most of the galaxies are at a
distance comparable to or less than that of the cluster then $\beta$ may be
much less than unity since many of the galaxies will be in front of the cluster
and not sheared at all or not far enough behind the cluster to receive the full
amount of shear.

\begin{figure}
\caption{Plotted is the mean tangential shear in a disk on the sky centered on
a model cluster vs. the redshift of the cluster. The different curves represent
different disk radii: from black to light gray the radii are $1'$, $2'$, $5'$,
$10'$, $20'$, $50'$.  The different plots are for different limiting
$b$ magnitudes as labeled.  Here we assume a Schechter luminosity function with
$\phi_*=0.014(h/{\rm Mpc})^3$, $\alpha=-0.97$, and $M_*=-19.5$ corresponding to
$b_J$ magnitudes. The model cluster has 1d velocity dispersion
$v_{\s1}=800$km/sec, core radius $\rc=250\,h^{-1}$kpc, and truncation radius
$\rt=3\,h^{-1}$Mpc.}
\end{figure}

	The shear around a given cluster is maximized a few core radii from the
center, while the maximal shear varies roughly proportional to $z$ until the
cluster distance approaches the depth of the survey.  One never finds large
shear too close to the cluster center and for more nearby clusters one must
look very far from the center to maximize the shear. Fig~1 illustrates that for
shallow surveys one is most sensitive to nearby structures.  Note that a disk
radius of $5'$ does a good job of maximizing the shear over a broad range of
cluster redshifts and limiting magnitudes. Wider area coverage yields a larger
signal only for $z\simlt0.2$ clusters.  Of course a large signal is of no use
unless one has sufficient galaxy numbers to detect it.  In fig~2 we see that
the available S/N is indeed significantly higher for deeper surveys but, with
large enough area coverage, can be much larger than unity even for very shallow
surveys. For extremely low redshift clusters one must survey very large areas
to obtain significant signal.  Yet even for $b<21$ one can in principle obtain
a significant signal from a $z=0.03$ cluster, like Coma, if one is able to
survey $\sim{1^\circ}^2$. Note however that this would require keeping
systematics well below the 1\% level.

\begin{figure}
\caption{Plotted is log$_{10}$S/N for measuring the amplitude of the shear for
clusters as in figure 1. The noise is assumed dominated by the finite number of
non-round galaxies in the sample assumed to have rms ellipticity 0.3.}
\end{figure}

	Given the low tolerance for systematic errors it is important to have a
good handle on how well one is determining the shear.  Besides simulations and
comparison with better (i.e. HST) data, one can also use an internal check of
one's data.  To do this take one's measured ellipticities and rotate their
position-angle by 45$^\circ$.  Then use one's favorite reconstruction technique
to estimate the surface density from the rotated data.  The surface density one
obtains should be consistent with zero up to the noise from the random galaxy
orientations and known measurement errors; and from effects due to the boundary
of one's sample. If not, one probably has discovered some systematic problems
with one's method.  The mathematics behind this is as follows.  One is trying
to estimate the shear tensor, $\gamma_{ab}$.  Such a 2-d symmetric traceless
tensor field can be decomposed into its scalar and pseudo-scalar parts
\begin{equation}
\gamma_{\rm s}=\nabla^{-2}\gamma_{ab,ab} \qquad
\gamma_{\rm p}=\nabla^{-2}\epsilon_{ac}\gamma_{ab,bc} \qquad
\epsilon_{ab}=\left(\begin{array}{cc} 0\ &          -1 \\
                                      1\ &\phantom{+}0 \end{array}\right),
\end{equation}
which is analogous to decomposing a vector into its curl and a curl-free
parts. For {\it weak} lensing $\gamma_{\rm s}$ is just proportional to the
weighted surface density while for gravitationally induced shear from
non-relativistic matter $\gamma_{\rm p}=0$ since the shear is derived from a
potential. Multiplying the shear tensor by $\epsilon_{ab}$ is the same as
rotating the position-angle of the shear by 45$^\circ$, so one obtains the
above result. The two components, $\gamma_{\rm s}$ and $\gamma_{\rm p}$, are so
similar that most sources of noise and error will contribute equally to both,
while the true signal will contribute only to $\gamma_{\rm s}$.  Thus it is
probably fair to believe one's results only to the extent that, on average,
$|\gamma_{\rm s}|>|\gamma_{\rm p}|$.  Kaiser and Tyson report that they have
used similar methods.

\begin{figure}
\caption{By rotating the measured ellipticities ({\it black}) by 45$^\circ$ and
then using these rotated ellipticities ({\it gray}) to reconstruct the surface
density one constructs a realization of the same size as the error in the
reconstructed surface density.}
\end{figure}

	At this writing the SDSS telescope is not yet operational and hence it
is difficult to know how it will perform in practice.  To address this issue
the authors have begun an observational program with a telescope at the SDSS
site, the ARC 3.5m telescope, using the Fermilab Drift Scan Camera (DSC) which
is similar, if much smaller, than the SDSS camera.  We have not yet reduced the
level of systematics to the point which would make the SDSS Northern survey
useful for weak lensing, but are confident that significant improvements will
be made.  The SDSS collaboration is in the process of comparing DSC data in
the Sloan colors and at the Sloan depth with deep HST WFPC-2 data.  This will
be extremely useful for gauging the accuracy of shear measurements that can be
expected from the SDSS.  If everything works well, the weak lensing data from
the SDSS northern survey will be one of its major achievements.  In any case we
certainly do expect that the deeper SDSS southern survey will yield useful
information from weak lensing studies.

\noindent $\bullet$ This work was supported by the DOE and NASA grant
\# NAG-5-2788.

{}


\begin{thebibliography}{}
\bibitem[\protect\citeauthoryear{Blandford {\it et al.}}{1991}]{BSBV}
R. Blandford, A. Saust, T. Brainerd, J. Villumsen (1991) {\it MNRAS},
                                                        {\bf 251}, pp.~600--627
\bibitem[\protect\citeauthoryear{Fahlman {\it et al.}}{1995}]{FKSW}
G. Fahlman, N. Kaiser, G. Squires, and D. Woods (1995) {\it Ap.J.}, {\bf 437},
                                                                     pp.~56--62
\bibitem[\protect\citeauthoryear{Kaiser}{1992}]{K92}
N. Kaiser (1992) {\it Ap.J.}, {\bf 388}, pp.~272--286
\bibitem[\protect\citeauthoryear{Kent}{1994}]{Kent94}
S. Kent (1994) {\it Ap. \& Space Sci.}, {\bf 217}, pp.~27--30
\bibitem[\protect\citeauthoryear{Miralda-Escud\'e}{1991}]{ME}
J. Miralda-Escud\'e (1991) {\it Ap.J.}, {\bf 380}, pp.~1--8
\bibitem[\protect\citeauthoryear{Mould {\it et al.}}{1994}]{MBVBSSK}
J. Mould {\it et al.} (1994) {\it MNRAS}, {\bf 271}, pp.~56--62
\bibitem[\protect\citeauthoryear{Smail {\it et al.}}{1994}]{SEF94}
I. Smail, R. Ellis, and M. Fitchett (1994) {\it MNRAS}, {\bf 270}, pp.~245--270
\bibitem[\protect\citeauthoryear{Szalay}{1995}]{Szalay}
A. Szalay (1995) private communication
\bibitem[\protect\citeauthoryear{Tyson {\it et al.}}{1990}]{TVW}
J. Tyson, F.Valdes, and R. Wenk (1990) {\it Ap.J. Lett}, {\bf 349}, pp.~L1-L4
\bibitem[\protect\citeauthoryear{Valdes {\it et al.}}{1983}]{VTJ}
F.Valdes, J. Tyson, and J. Jarvis (1983) {\it Ap.J.}, {\bf 271}, pp.~431--441
\bibitem[\protect\citeauthoryear{Gould and Villumsen }{1994}]{VG94}
A. Gould  and J. Villumsen (1994) , {\it Ap.J. Lett.}, {\bf 428}, pp.~L45--L48

\end{thebibliography}
\end{document}